\documentstyle[astrobib,psfig]{mn-ab}

\title[Structure in the Ly$\alpha$ forest II] {Large-scale structure
in the Lyman-$\alpha$ forest II: analysis of a group of ten QSOs}

\author[J. Liske et al.]{J.~Liske,$^1$\thanks{E-mail: 
	jol@phys.unsw.edu.au} J.~K.~Webb,$^1$ G.~M.~Williger,$^2$
	A.~Fern\'andez-Soto$^1$
	\newauthor
	and R.~F.~Carswell$^3$\\ 
	$^1$School of Physics, University of New South Wales, Sydney 2052,
	Australia\\ 
	$^2$NASA Goddard Space Flight Center, Greenbelt, Maryland 20771, USA\\
	$^3$Institute of Astronomy, Madingley Road, Cambridge CB3 0HA}

\date{Accepted
...... Received .....}

\pagerange{\pageref{firstpage}--\pageref{lastpage}}

\pubyear{1999}

\newcommand{\lya}{\mbox{Ly$\alpha$}}
\newcommand{\be}{\begin{equation}}
\newcommand{\ee}{\end{equation}}
\newcommand{\eref}[1]{(\ref{#1})}

\begin{document}

\label{firstpage}
\maketitle

\begin{abstract}
The spatial distribution of \lya\ forest absorption systems towards
ten QSOs has been analysed to search for large-scale structure over
the redshift range $2.2 < z < 3.4$. The QSOs form a closely spaced
group on the sky and are concentrated within a $1$~deg$^2$ field.  We
have employed a technique based on the first and second moments of the
transmission probability density function which is capable of
identifying and assessing the significance of regions of over- or
underdense \lya\ absorption. We find evidence for large-scale
structure in the distribution of \lya\ forest absorption at the $> 99$
per cent confidence level. In individual spectra we find overdense
\lya\ absorption on scales of up to $1200$~km~s$^{-1}$. There is also
strong evidence for correlated absorption across line of sight pairs
separated by $< 3$~h$^{-1}$ proper Mpc ($q_0 = 0.5$). For larger
separations the cross-correlation signal becomes progressively less
significant.
\end{abstract}


\section{Introduction}
The \lya\ forest seen in the spectra of distant QSOs may constitute a
substantial fraction of the baryonic content of the Universe
\cite{Rauch95} and its evolution can be traced over most of the
history of the Universe.  Every QSO absorption spectrum provides us
with a representative, albeit one-dimensional sample of the baryonic
matter distribution, unaffected by any luminosity bias. The intimate
relationship of the absorbing gas with the thermal, chemical, and
dynamical history of the Universe makes the \lya\ forest a versatile,
but not always user-friendly tool in the study of cosmology.

The observational and theoretical advances of the past decade have
significantly altered our understanding of the \lya\ forest. The
advent of HST's UV spectroscopic capabilities led to the first
detailed analyses of the low redshift \lya\ forest (\citeNP{Morris91};
\citeNP{Bahcall91}). Subsequently, using multislit spectroscopy and
deep imaging of galaxies in the fields of QSOs, significant numbers of
coincidences between the redshifts of absorption lines and galaxies
were found at $z \la 1$. Several groups determined that galaxies have
absorption cross-sections of $\sim 200$~h$^{-1}$~kpc
(\citeNP{Lanzetta95}; \citeNP{LeBrun96}; \citeNP{Bowen96}). Moreover,
\citeN{Lanzetta95}, \citeN{Chen98}, and \citeN{Tripp98} found an
anti-correlation of the \lya\ rest equivalent width, $W$, with the
distance of the absorbing galaxy to the line of sight of the QSO as
well as a slight correlation with luminosity \cite{Chen98}. Thus
\citeN{Chen98} confirmed earlier results that the extended gaseous
halos of galaxies are directly responsible for a significant fraction
of the \lya\ forest at $z \la 1$. However, there is evidence that this
may only be true for stronger lines ($W \ga 0.3$~\AA) and that the
weaker absorption lines trace the large-scale gaseous structures in
which galaxies are presumably embedded (\citeNP{LeBrun98};
\citeNP{Tripp98}).

At high redshift, the discovery of measurable amounts of C~{\small IV}
associated with 75 per cent of all \lya\ absorbers with coloumn
density $N($H~{\small I}$) > 10^{14.5}$~cm$^{-2}$ (\citeNP{Cowie95};
\citeNP{Songaila96}) in high-quality spectra obtained with the HIRES
spectrograph on the Keck 10-m telescope has challenged the original
notion of the \lya\ forest arising in pristine, primordial
gas. However, \citeN{Lu98} found almost no associated C~{\small IV}
absorption for systems with $N($H~{\small I}$) < 10^{14}$~cm$^{-2}$
and derived $[{\rm C}/{\rm H}] < -3.5$, about a factor of ten smaller
than inferred by \citeN{Songaila96} (see also \citeNP{Dave98}),
suggesting a sharp drop in the metallicity of the \lya\ forest at
$N($H~{\small I}$) \approx 10^{14}$~cm$^{-2}$. Nevertheless, the
metallicity of the higher column density systems raises the
possibility of an association of these systems with galaxies at high
redshift. Using the same C~{\small IV} lines to resolve the structure
of the corresponding blended \lya\ lines, \citeN{Fernandez96} have
shown that the clustering properties of these systems are consistent
with the clustering of present-day galaxies if the correlation length
of galaxies is allowed to evolve rapidly with redshift ($\epsilon
\approx 2.4$).

On the theoretical side, models have progressed from the early
pressure-confined (\citeNP{SYBT}; \citeNP{Ostriker83};
\citeNP{Ikeuchi86}; \citeNP{Williger92}) and dark matter mini-halo
(\citeNP{Rees86}; \citeNP{Ikeuchi86b}) scenarios to placing the \lya\
forest fully within the context of the theory of CDM dominated,
hierarchical structure formation. Both semi-analytical
(\citeNP{Petitjean95}; \citeNP{Bi97}; \citeNP{Hui97};
\citeNP{Gnedin98b}) and full hydrodynamical numerical simulations
(\citeNP{Cen94}; \citeNP{Zhang95}; \citeNP{Miralda96};
\citeNP{Hernquist96}; \citeNP{Wadsley97}; \citeNP{Theuns98b}) of
cosmological structure formation, which include the effects of
gravity, photoionisation, gas dynamics, and radiative cooling, have
shown the \lya\ forest to arise as a natural by-product in the
fluctuating but continuous medium which forms by gravitational growth
from initial density perturbations. The simulations are able to match
many of the observed properties of the \lya\ absorption to within
reasonable accuracy (\citeNP{Miralda96}; \citeNP{Zhang97};
\citeNP{Dave97}; \citeNP{Dave98b}; \citeNP{Muecket96};
\citeNP{Riediger98}; \citeNP{Theuns98}). A common feature of all the
simulations is that the absorbing structures exhibit a variety of
geometries, ranging from low density, sheet-like and filamentary
structures to the more dense and more spherical regions where the
filaments interconnect and where, presumably, galaxies form. The
dividing line between these different geometries lies in the range
$10^{14}$~cm$^{-2} \la N \la 10^{15}$~cm$^{-2}$ (\citeNP{CenSim97};
\citeNP{Zhang98}).

Thus a consistent picture may be emerging: absorption lines with
$N \ga 10^{14.5}$~cm$^{-2}$ are closely associated with the large,
spherical outer regions of galaxies, cluster strongly on small
velocity scales along the line of sight \cite{Fernandez96}, and have
been contaminated with metals by supernovae from a postulated
Population III (e.g. \citeNP{Miralda97}) or by galaxy mergers
(e.g. \citeNP{Gnedin98}); whereas lower column density lines trace the
interconnecting, filamentary structures of the intergalactic medium.

Whatever the case may be, it seems likely that the large-scale
distribution of the \lya\ absorption holds important clues to its
origin. \lya\ absorbers are probably fair tracers of the large-scale
cosmic density field and should thus be able to constrain structure
formation models. To investigate further the connection of the \lya\
forest with cosmological structure we study in this paper its
large-scale structure both in velocity and real space by considering a
close group of ten QSOs. The QSOs are contained within a $\sim
1$~deg$^2$ field so that the \lya\ forest is probed on Mpc scales.

There have been many studies of the clustering properties of the \lya\
forest. \citeN{Webb87a} was the first to report a weak signal in the
two-point correlation function of fitted absorption lines on scales of
$\la 100$~km~s$^{-1}$. This result was later confirmed
(\citeNP{Ostriker88}; \citeNP{Chernomordik95}; \citeNP{Cristiani95b};
\citeNP{Kulkarni96}; \citeNP{Cristiani97}; \citeNP{Khare97}), and some
investigators found considerably stronger signals (\citeNP{Ulmer96};
\citeNP{Fernandez96}). There have also been detections on larger
scales. \citeN{Meiksin95} used a nearest neighbour statistic to derive
correlations on scales of $0.5$--$3$~h$^{-1}$~Mpc. \citeN{Pando96}
employed the discrete wavelet transform to demonstrate the existence
of clusters on scales of $10$--$20$~h$^{-1}$~Mpc in the \lya\ forest
at a significance level of $2$--$4\sigma$ and that the number density
of these clusters decreases with increasing redshift. More recently,
they observed non-Gaussian behaviour of \lya\ forest lines on scales
of $5$--$10$~h$^{-1}$~Mpc at a confidence level larger than 95 per
cent \cite{Pando98b}. \citeN{Mo92} even reported $60$ and
$130$~h$^{-1}$~Mpc as characteristic scales of the \lya\ forest. All
of these studies however are based on analyses of individual lines of
sight.

By comparing the absorption characteristics in two (or more) distinct,
but spatially close lines of sight, it is possible to investigate
directly the real space properties of \lya\ systems on various
scales. The multiple images of gravitationally lensed QSOs have been
used to establish firm lower limits on the sizes of \lya\
clouds. \citeN{Smette95} found that the \lya\ forests of the two
images of HE~1104--1805 (separation $3.0$ arcsec) are virtually
identical and inferred a $2\sigma$ lower limit of $50$~h$^{-1}$~kpc on
the diameter of \lya\ clouds, assumed to be spherical, at $z =
2.3$. More information on the sizes of \lya\ absorbers has been gained
from the studies of close QSO pairs with separations of several arcsec
to $\sim 7$~arcmin. All of the most recent analyses have concluded
that \lya\ absorbers have diameters of a few hundred kpc
(\citeNP{Fang96}; \citeNP{Dinshaw97}; \citeNP{Dinshaw98};
\citeNP{Petitjean98}; \citeNP{DOdorico98}) and \citeN{Crotts98} found
that correlations among neighbouring lines of sight persist for
separations up to $0.5$--$0.8$~h$^{-1}$~Mpc for lines with $W >
0.4$~\AA. A tentative detection of increasing cloud size with
decreasing redshift (as would be expected for absorbers expanding with
the Hubble flow) was reported by \citeN{Fang96} and
\citeN{Dinshaw98}. The possibility remains however, that the
coincidences of absorption lines are due to spatial clustering of
absorbers and that the inferred `sizes' in fact are an indication of
their correlation length (\citeNP{Dinshaw98}; \citeNP{CenSim97}) as
may be evidenced by the correlation of the estimated `size' with line
of sight separation found by \citeN{Fang96}.

Analyses to determine the shape of absorbers, as proposed e.g. by
\citeN{Charlton95}, have so far been inconclusive. Several authors
agree that the current data are incompatible with uniform-sized
spherical clouds but are unable to decisively distinguish between
spherical clouds with a distribution of sizes, flattened disks, or
filaments and sheets (\citeNP{Fang96}; \citeNP{Dinshaw97};
\citeNP{DOdorico98}). However, from the analysis of a QSO triplet
\citeN{Crotts98} found evidence that lines with $W > 0.4$~\AA\ arise
in sheets.

For still larger line of sight separations most work has concentrated
on metal absorption. \citeN{Williger96}, e.g., used a group of 25 QSOs
contained within a $\sim 1$~deg$^2$ region to identify structure in
C~{\small IV} absorption on the scale of $15$--$35$~h$^{-1}$~Mpc. Within
this group, a subset of ten QSOs is suitable for studying the
large-scale structure of the \lya\ forest. A cross-correlation
analysis of these data, performed by \citeN{Williger99}, revealed a
$3.7\sigma$ excess of line-pairs at velocity splittings $50 < \Delta v <
100$~km~s$^{-1}$. Thus \citeANP{Williger99} concluded that the \lya\
forest seems to exhibit structure on the scale of $\sim
10$~h$^{-1}$~Mpc in the plane of the sky. However there was no excess
at smaller velocity splittings and no dependence of the signal on
angular separation was found.

Finally, the largest line of sight separations were investigated by
\citeN{Elowitz95}, who studied a group of four QSOs, projected within
$2\fdg8$ on the sky. Probing scales of $\sim 30$~h$^{-1}$~Mpc in the
plane of the sky, no significant cross-correlation signal was found
out to a velocity separation of $10^4$~km~s$^{-1}$.

From the results outlined above it appears that the \lya\ forest shows
significant correlations across lines of sight at all but the largest
scales probed. In this paper we identify more precisely the upper
limit of these correlations. To this end we re-analyse the group of
ten QSOs of \citeANP{Williger99} in the South Galactic Pole
region. We employ a novel technique which is based on the statistics
of the transmitted flux rather than the statistics of fitted
absorption lines \cite{Liske98b} (hereafter paper I). The South
Galactic Cap region has one of the highest known QSO densities in the
sky and the dataset is the most useful for large-scale structure
studies of the \lya\ forest published so far. We find evidence for a
transition from strong correlations for proper line of sight
separations\footnote{We use $q_0 = 0.5$, $\Lambda = 0$ and $H_0 =
100$~h~km~s$^{-1}$~Mpc$^{-1}$ throughout this paper.} $<
3$~h$^{-1}$~Mpc to a vanishing correlation for line of sight
separations $> 6$~h$^{-1}$~Mpc.

The rest of the paper is organised as follows: in Section \ref{data}
we briefly describe the data. Section \ref{analysis} outlines the
method of analysis and its main advantages over a `traditional'
two-point correlation function analysis. Our results are presented in
Section \ref{results} and discussed in Section \ref{conclusions}.

\section{The Data} \label{data}
The data were gathered as part of a larger survey designed to reveal
the large-scale clustering properties of C~{\small IV} systems
\cite{Williger96}. The observations were made on the CTIO 4-m
telescope using the Argus multifibre spectrograph. The instrumental
resolution was $\sim 2$~\AA\ and the signal-to-noise ratio per pixel
reached up to 40 per 1~\AA\ pixel. For a complete description of the
observations and the reduction process we refer the reader to
\citeN{Williger96}.

\begin{figure}
\psfig{file=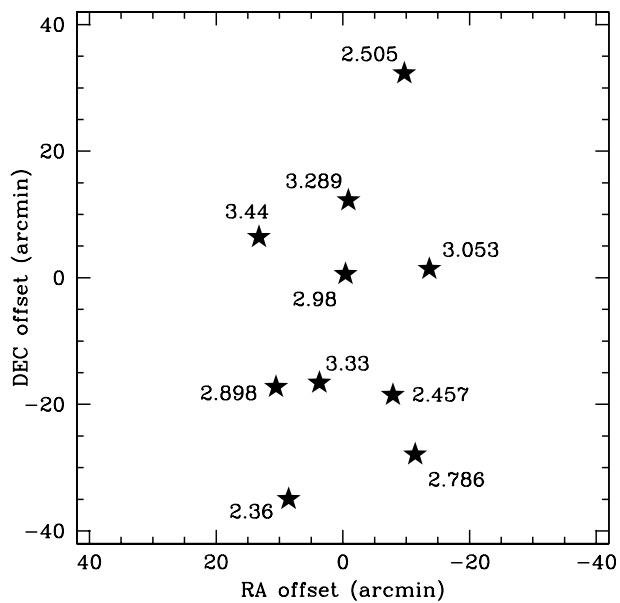,width=\columnwidth,silent=}
\caption{Distribution of QSOs in the sky. The field is centered on
$\alpha = 00^{{\rm h}} 42^{{\rm m}} 10^{{\rm s}}$ and $\delta =
-26\degr 40\arcmin$ (B1950). Stars mark the positions of the QSOs
listed in Table \ref{qsotab} and emission redshifts are indicated.}
\label{qsos}
\end{figure}

Of the original sample of $25$ QSO spectra only $14$ cover any part of
the region between the \lya\ and Ly$\beta$ emission lines. In order to
avoid the proximity effect we exclude from the analysis those parts of
the spectra which lie within 3000~km~s$^{-1}$ of the \lya\ emission
line. This leaves us with $11$ spectra. We have excluded an additional
spectrum from the analysis (Q0043--2606) because of its low S/N and
small wavelength coverage. We are thus left with the same sample as
used by \citeN{Williger99}. We list these QSOs, their positions and
redshifts in Table \ref{qsotab} and show their distribution in the sky
in Fig.~\ref{qsos}. The data cover $2.17 < z < 3.40$.
\begin{table}
\caption{QSOs analysed.}
\label{qsotab}
\begin{tabular}{cccc}
\hline
Object &  $\alpha_{1950}$ & $\delta_{1950}$ & $z_{em}$ \\ 
 & $\!\!\! h \  m \quad \  s$ & $\degr \quad \arcmin \quad \arcsec$ & \\
\hline
Q0041--2638	 & $00\; 41\; 15.19$ & $-26\; 38\; 35.9$ & $3.053$ \\
Q0041--2707	 & $00\; 41\; 24.38$ & $-27\; 07\; 54.3$ & $2.786$ \\
Q0041--2607	 & $00\; 41\; 31.11$ & $-26\; 07\; 41.7$ & $2.505$ \\
Q0041--2658	 & $00\; 41\; 38.38$ & $-26\; 58\; 30.0$ & $2.457$ \\
Q0042--2627	 & $00\; 42\; 06.42$ & $-26\; 27\; 45.3$ & $3.289$ \\
Q0042--2639    	 & $00\; 42\; 08.20$ & $-26\; 39\; 25.0$ & $2.98$ \\
Q0042--2656	 & $00\; 42\; 24.89$ & $-26\; 56\; 34.4$ & $3.33$ \\
Q0042--2714	 & $00\; 42\; 44.12$ & $-27\; 14\; 56.6$ & $2.36$ \\
Q0042--2657	 & $00\; 42\; 52.29$ & $-26\; 57\; 15.3$ & $2.898$ \\
Q0043--2633	 & $00\; 43\; 03.10$ & $-26\; 33\; 33.6$ & $3.44$ \\
\hline
\end{tabular}
\end{table}
The angular separations range from $6.1$ to $69.2$ arcmin or $1.4$ to
$16.3$~h$^{-1}$~proper~Mpc and the emission redshifts range from
$2.36$ to $3.44$.

\section{Analysis} \label{analysis}
The method of analysis used in this work is described in detail in
paper I. Our technique does not rely on the statistics of individual
absorption lines but rather identifies local over- or underdensities
of \lya\ absorption using the statistics of the transmitted flux. In
order to identify large-scale features in a normalised spectrum of the
\lya\ forest we convolve the spectrum with a smoothing function. In
fact, we smooth the spectrum on all possible scales. On the smallest
possible scale ($= 1$ pixel) the spectrum remains essentially
unchanged whereas on the largest possible scale ($=$ number of pixels
in the spectrum) the whole spectrum is compressed into a single number
akin to $1 - D_A$, where $D_A$ is the flux deficit parameter
\cite{Oke82}. When plotted in the wavelength--smoothing scale plane,
this procedure results in the `transmission triangle' of the
spectrum. The base of the triangle is the original spectrum itself and
the tip is the single value which results from smoothing the spectrum
on the largest possible scale. Thus a cluster of absorption lines will
be enhanced by filtering out the `high-frequency noise' of the
individual absorption lines. We use a Gaussian as the smoothing
function and denote the smoothed spectrum by $G(\lambda, \sigma_{\rm
s})$.

We are interested in identifying local fluctuations around the mean
\lya\ transmission and in assessing their statistical significance. To
this end we have calculated both the expected mean, $\langle G
\rangle$, and variance, $\sigma_G^2$, of the \lya\ transmission as
functions of wavelength and smoothing scale in paper I. The
calculations are based on the simple null-hypothesis that any \lya\
forest spectrum can be described by a collection of individual
absorption lines whose parameters are uncorrelated. Thus, in
particular, the null-hypothesis presumes an unclustered \lya\
forest. The derivation also uses the observationally determined
functional form of the distribution of the absorption line parameters,
for redshift $z$, column density $N$, and Doppler parameter $b$,
\be \label{Vd}
\eta(z, N, b) \propto (1 + z)^\gamma N^{-\beta} \exp\left[\frac{-(b - \mu_b)^2}
{2 \sigma_b^2}\right]
\ee
for $b > b_{\rm cut}$ (\citeNP{Carswell84}; \citeNP{Lu96};
\citeNP{Kim97}; \citeNP{Kirkman97}) and the results are given in terms
of its parameters:
\be \label{mG}
\langle G \rangle (\lambda, \sigma_{\rm s}) = 
{\rm e}^{-B (\frac{\lambda}{\lambda_\alpha})^{\gamma+1}},
\ee
and
\be \label{sigG}
\sigma_G^2(\lambda, \sigma_{\rm s}) = \frac{\sigma_n^2(\lambda)}
{2\sqrt{\pi} \; \sigma_{\rm s}/ps} + \frac{\sigma_{{\rm e}^{-\tau}}^2(\lambda)}
{\sqrt{2\frac{\sigma_{\rm s}^2 + \sigma_{\rm LSF}^2}{q^2(\lambda)} + 1}},
\ee
where
\be \label{sigmaetau}
\sigma_{{\rm e}^{-\tau}}^2 = {\rm e}^{-2^{\beta-1} B 
(\frac{\lambda}{\lambda_\alpha})^{\gamma+1}} - 
{\rm e}^{-2 B (\frac{\lambda}{\lambda_\alpha})^{\gamma+1}}.
\ee
$\sigma_n$ denotes the noise of a spectrum, $\sigma_{\rm LSF}$ the
width of the line spread function, $ps$ the pixel size in \AA, and
$\lambda_{\alpha} = 1215.67$~\AA. $q$ is the intrinsic width of the
auto-covariance function of a spectrum (approximated as a Gaussian,
see also paper I). In paper I we tested the analytic expressions above
against extensive numerical simulations and found excellent
agreement. Given that the Gunn-Peterson optical depth is limited to
$\tau_{\rm GP} \la 0.04$ \cite{Webb92} and given the number density of
metal lines, we can safely neglect the effects of both of these
potential contributors to $\tau_{\rm eff}$ in equation \eref{mG}.

A potential source of error in the method used here is its sensitivity
to the adopted continuum fit. A small error in the zeroth or first
order of the fit introduces an arbitrary offset from the expected mean
transmission. However, this is easily dealt with by determining the
normalisation of the mean optical depth, $B$, for each spectrum
individually:
\be \label{norm}
B = -\left(\frac{\lambda_\alpha}{\lambda_{\rm c}}\right)^{\gamma + 1} 
\ln[G(\lambda_{\rm c}, \sigma_{\rm s, max})], 
\ee
where $\sigma_{\rm s, max}$ denotes the largest possible smoothing
scale and $\lambda_{\rm c}$ is the central wavelength of the spectrum. 
Thus we fix the normalisation for each spectrum at the tip of its 
transmission triangle.

Before we can evaluate expressions \eref{mG} and \eref{sigG} we need
the values of several other parameters: $\gamma$ was determined by
\citeN{Williger99} for the present data to be $2.5$ and we take
$\beta$ from recent high resolution studies as $1.5$ (\citeNP{Hu95};
\citeNP{Lu96}; \citeNP{Kim97}; \citeNP{Kirkman97}). As in paper I we
determined the width of the auto-covariance function of a `perfect'
spectrum (i.e.~before it passes through the instrument), $q$, from
simulations. We simulated each spectrum of the dataset $100$ times in
accordance with the null-hypothesis (see also paper I), using the
$\gamma$ and $\beta$ values as above and $\mu_b = 30$~km~s$^{-1}$,
$\sigma_b = 8$~km~s$^{-1}$ and $b_{\rm cut} = 18$~km~s$^{-1}$
(\citeNP{Hu95}; \citeNP{Lu96}; \citeNP{Kim97}; \citeNP{Kirkman97}).
The simulated data have the same resolution as the real data but a
constant (conservative) S/N of $20$. The simulations are normalised to
give the same mean effective optical depth as the real data.

With the values of all the parameters in place, we can transform a
given transmission triangle into a `reduced' transmission triangle by
\be \label{rg}
RG(\lambda, \sigma_{\rm s}) = \frac{G - \langle G \rangle}{\sigma_G}.
\ee
The reduced transmission triangle shows the residual fluctuations of
the \lya\ transmission around its mean in terms of their statistical
significance.

There are several advantages of the method outlined above over a more
`traditional' approach involving absorption line fitting and
construction of the two-point correlation function. Most importantly,
our technique does not require arduous line-fitting and side-steps all
difficulties arising from line-blending \cite{Fernandez96}. In paper I
we demonstrated that for simulated data our technique is significantly
more sensitive to the presence of large-scale structure than a
two-point correlation function analysis, especially at intermediate
resolution. Moreover, numerical simulations of cosmological structure
formation (\citeNP{Cen94}; \citeNP{Miralda96}; \citeNP{Hernquist96};
\citeNP{Zhang95}; \citeNP{Wadsley97}; \citeNP{Theuns98b}) indicate
that at least the low column density forest arises in a fluctuating
but continuous medium so that a given absorption line may not
correspond to a well-defined individual `cloud'. Thus a method of
analysis based on the statistics of the transmitted flux itself is
most appropriate and intuitive.

A high signal-to-noise ratio in the QSO spectra is not of great
importance for the analysis itself (as the noise is quickly smoothed
over, see equation \ref{sigG}), but it is important for reliable
continuum fitting.  Uncertainties in the placement of the continuum
have already been dealt with above (equation \ref{norm}). The second
source of uncertainty are the values of the parameters of equation
\eref{Vd} and we will discuss the pertaining effects in Sections
\ref{lm} and \ref{dlos}.

\section{Results} \label{results}
\subsection{Single lines of sight} \label {slos}
\begin{figure*}
\psfig{file=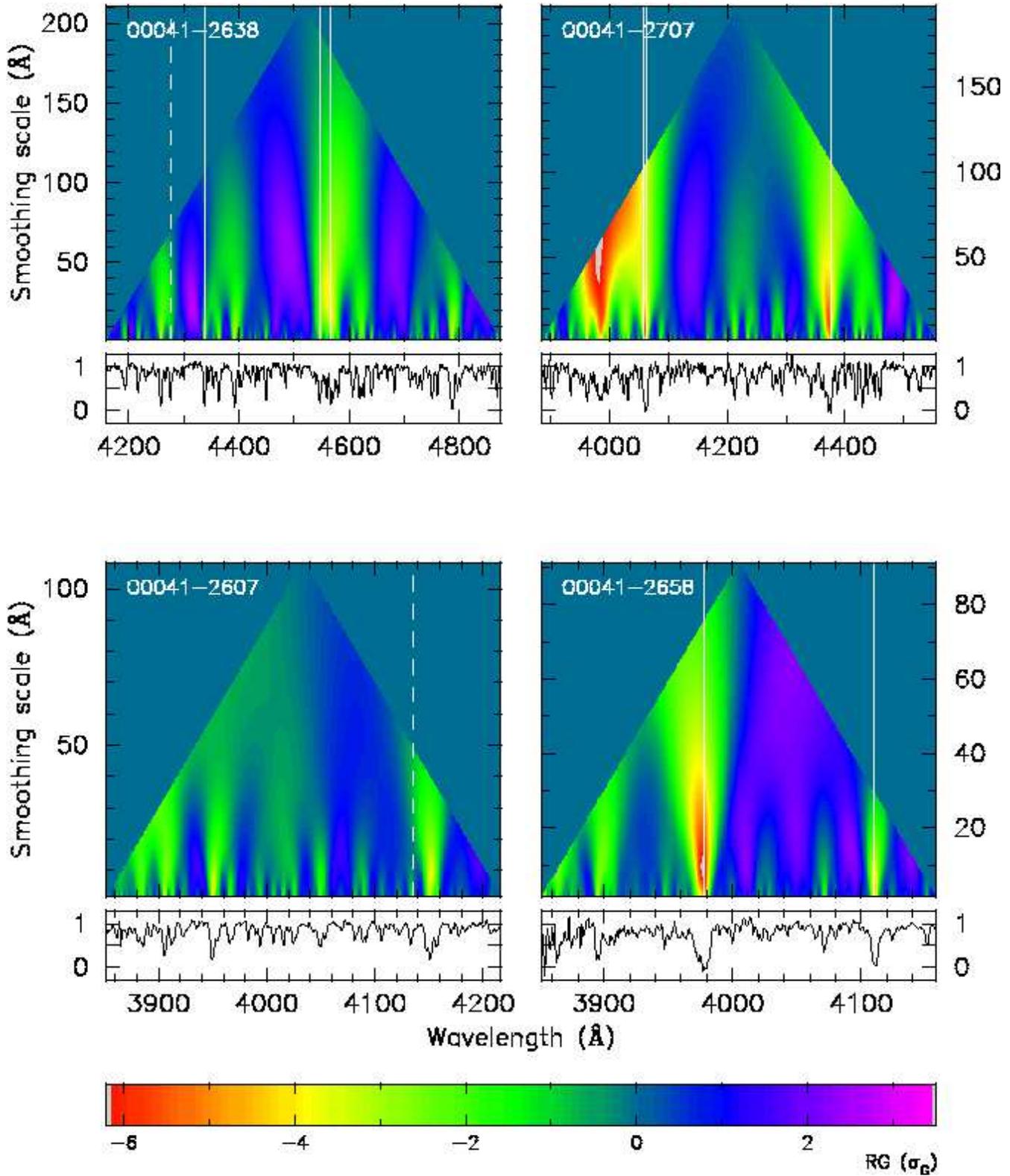,width=\textwidth,silent=}
\caption{Reduced transmission triangles and normalised spectra of four
of the QSOs listed in Table \ref{qsotab}. The smoothing scale on the
vertical axis is the FWHM of the smoothing Gaussian. Regions of
overdense absorption appear yellow and red, underdense regions appear
blue and purple. The vertical lines indicate the observed \lya\
wavelengths of known metal absorption systems. Dashed lines mark the
positions of metal systems that do {\em not} have an associated \lya\
local minimum (see text).}
\label{mosaic}
\end{figure*}
In Fig.~\ref{mosaic} we show the result of the analysis described
above for four of the QSOs listed in Table \ref{qsotab} which show 
the most significant features as described below. The vertical
lines show the \lya\ positions of known metal systems which were
primarily taken from \citeN{Williger96} (their Table 3). A search
using NED\footnote{The NASA/IPAC Extragalactic Database (NED) is
operated by the Jet Propulsion Laboratory, California Institute of
Technology, under contract with the National Aeronautics and Space
Administration.} uncovered only three additional systems, all towards
Q0042--2627 \cite{York91}. Not surprisingly we see overdense \lya\
absorption at the positions of {\em all} metal systems. We take this
as an indication that our method correctly identifies overdensities.

Fig.~\ref{mosaic} clearly demonstrates the presence of structures on
scales as large as many hundred km~s$^{-1}$. In all there are seven
features which are significant at the more than $-4\sigma_G$ level. We
now briefly describe these quite significant detections:

{\bf Q0041--2707:} This spectrum shows three very significant
overdensities in \lya. The first, $-6.2\sigma_G$ at $(\lambda, {\rm
FWHM}_{\rm s}) = (3983$~\AA, $4179$~km~s$^{-1})$, is caused by an
unsaturated cluster of lines which has no associated metal system.
The second very significant ($-6.0\sigma_G$) overdensity in this
spectrum lies at $(4062$~\AA, $498$~km~s$^{-1})$ and is caused by a
saturated blend. It seems to be associated with either or both of two
metal systems detected in C~{\small IV} and Si~{\small IV}, Si~{\small
II} and C~{\small IV} respectively. The third significant system is
again associated with metals, is significant at the $-5.6\sigma_G$
level, and lies at $(4373$~\AA, $745$~km~s$^{-1})$.

{\bf Q0041--2658:} The most significant overdensity in the entire
sample is a $-6.3\sigma_G$ overdensity at $(3977$~\AA,
$686$~km~s$^{-1})$, caused by a broad, clearly blended and saturated
feature which is associated with C~{\small II}, Si~{\small IV},
C~{\small IV} and Al~{\small II} absorption. There is also a
$-4.3\sigma_G$ overdensity at $(4111$~\AA, $278$~km~s$^{-1})$, again
associated with metals (S~{\small II} and C~{\small IV}).

{\bf Q0042--2627:} A $-4.5\sigma_G$ overdensity at $(4522$~\AA,
$1071$~km~s$^{-1})$, with no associated metals, caused by a broad
cluster of lines.

{\bf Q0043--2633:} A $-5.2\sigma_G$ overdensity at $(4641$~\AA,
$778$~km~s$^{-1})$, with no associated metals, caused by a broad,
saturated feature.

\subsection{Local minima} \label{lm}
The probability density function (pdf) of $G$ is inherently
non-Gaussian, such that e.g. the probability of $G$ lying within
$3\sigma_G$ of the mean is smaller than $0.9973$. In fact, the pdf of
$G$ is a function of smoothing scale with the Central Limit Theorem
guaranteeing a Normal distribution at very large smoothing scales. (We
have attempted to continually remind the reader of this point by the
persistent use of the notation `$\sigma_G$'.) In light of this
difficulty it is necessary to estimate reliable significance levels by
using simulated data.

To further demonstrate and quantify the significance of the
overdensities presented in the previous section we investigate the
statistics of local minima in the reduced transmission triangles by
comparing them with the statistics of local minima derived from
simulated data. We simply define a `local minimum' (LM) as any pixel
in a reduced transmission triangle with
\be 
RG(\lambda, \sigma_{\rm s}) < -2.0\sigma_G
\ee
and where all surrounding pixels have larger values. If there is more
than one LM in the same wavelength bin (but at different smoothing
scales) then we delete the less significant one, since we do not want
to count the same structure more than once. Given the resolution of
the data it is evident that unclustered absorption lines will produce
LM on the smallest smoothing scale ($= 1.5$~\AA). We may anticipate
that the total number and distribution of these LM depend sensitively
on the parameters of the underlying line distribution \eref{Vd}. Thus
we exclude all LM on the smallest smoothing scale from all further
analyses in order to reduce the model-dependence of our results.

Note that with this definition, almost all metal lines have a \lya\ LM
within $300$~km~s$^{-1}$ (vertical lines in Fig.~\ref{mosaic}).

As mentioned in Section \ref{analysis}, we have simulated $100$
datasets ($= 1000$ spectra), essentially by putting down Voigt
profiles with parameters drawn from distribution \eref{Vd} (see paper
I for the exact technique). We then applied the same procedures as
outlined above: first we constructed the reduced transmission
triangles for all simulated spectra and then we determined the LM.

\begin{figure*}
\psfig{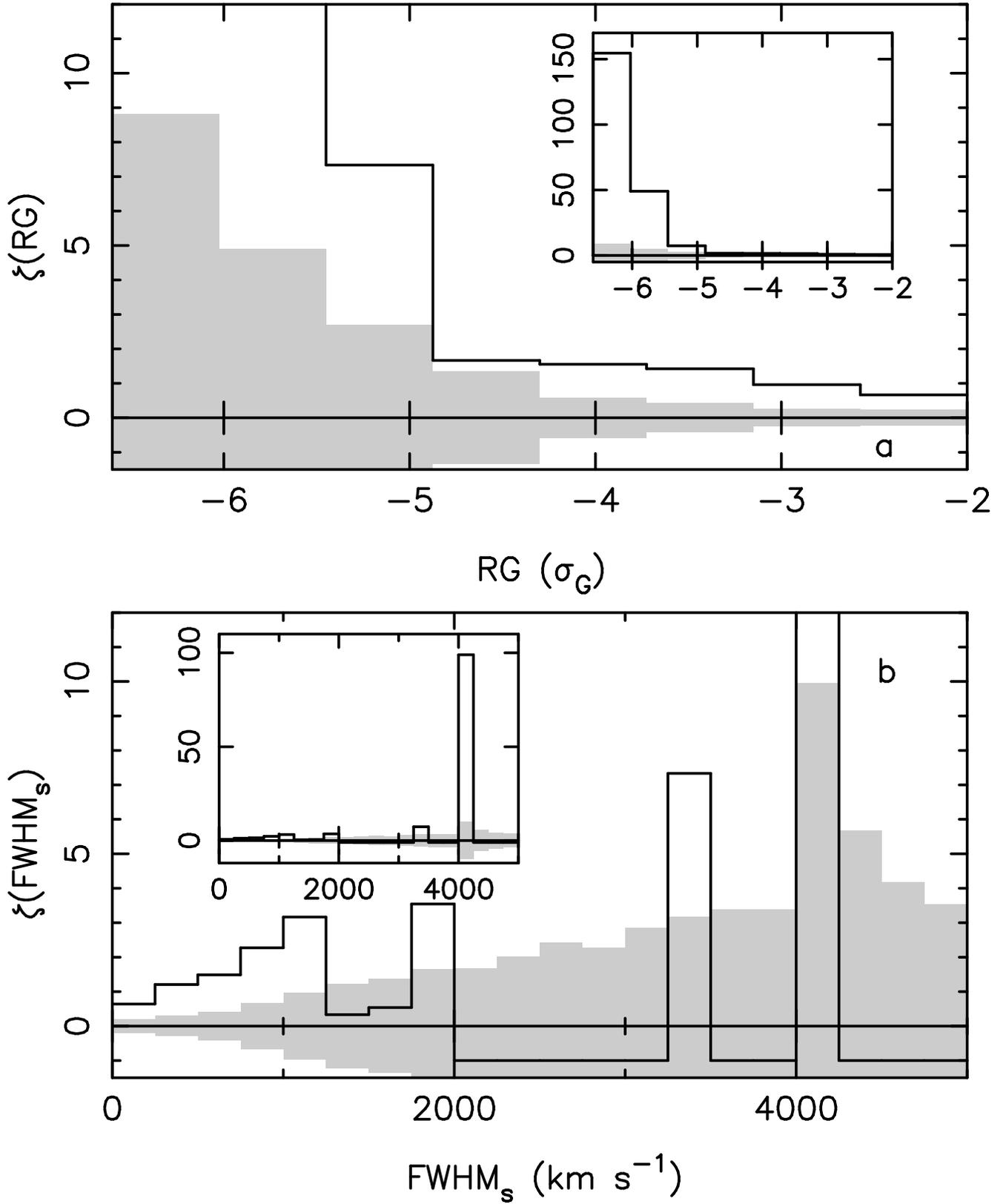}
\caption{(a) Excess of observed number of local minima in reduced
transmission triangles over the expected number ($=$ mean of $100$
simulations) as a function of the significance level of the local
minima (see equation \ref{zeta}). $\pm1$~rms levels as derived from
the simulations are indicated by the grey shaded areas. (b) Same as
(a) as a function of smoothing scale (FWHM of smoothing Gaussian). The
insets show the same plots but with expanded $\zeta$-scales.}
\label{lm_s}
\end{figure*}

In the real data we found $103$ LM whereas the simulated data yielded
on average $51$ LM with an rms dispersion of $7.6$. The maximum number
of LM found in the $100$ simulated datasets was $70$. Thus there is an
excess of the total number of LM over the expected number at the $>
99$ per cent confidence level. In Fig.~\ref{lm_s} we plot
\be \label{zeta}
\zeta(x) = \frac{N_{\rm obs}(x)}{N_{\rm exp}(x)} - 1,
\ee
where $N_{\rm obs}$ ($N_{\rm exp}$) is the observed (expected, as
derived from the simulations) number of LM and $x = RG$ (significance
level of LM in units of $\sigma_G$) or $x = {\rm FWHM}_{\rm s}$ (FWHM
of smoothing Gaussian).

From panel (a) we can see that there is a tendency for the excess to
be more significant for the stronger LM. For the leftmost bin we found
two LM in the data but {\em none} in the $100$ simulated
datasets. Extrapolating from lower significance levels we find $N_{\rm
exp} = 0.013$ for this bin, which is the value used in panel (a). A
Kolmogorov--Smirnov test indicates that the distributions $N_{\rm
obs}(RG)$ and $N_{\rm exp}(RG)$ disagree at the $95.6$ per cent
confidence level. The excess is strengthened further and the
confidence level is increased to $> 99$ per cent by excluding all LM
with smoothing scales $< 250$~km~s$^{-1}$ thus showing that the excess
is dominated by the larger scale LM. In smoothing scale-space we also
observe an excess which occurs on scales of up to $1200$~km~s$^{-1}$
(panel b). By excluding LM with $RG > -4\sigma_G$, the excess on
scales $< 250$~km~s$^{-1}$ disappears but it persists on larger scales
thus showing that it is dominated by the more significant LM.

Q0041--2707 and Q0041--2658 exhibit the most significant structures,
as can be seen in Fig.~\ref{mosaic} (and as discussed in Section
\ref{slos}). If these two objects are removed from the sample the
excess of LM is slightly reduced but remains significant at the
$6.1\sigma$ level. The effect also persists if we exclude
non-significant substructure from the analysis by deleting all LM
which lie within the subtriangle defined by another more significant
LM on larger smoothing scales. The removal of all LM that lie within
$300$~km~s$^{-1}$ of known metal lines again slightly weakens the
excess of LM but does not remove the effect.

If we have used incorrect values for the observational parameters of
equation \eref{Vd} then we have over- or underestimated either the
absolute value of the transmission fluctuations or their statistical
significance or both. This may affect both the normalisation and the
shape of the distributions $N_{\rm obs}(x)$. The exclusion of LM on
the smallest smoothing scale from the analysis was designed to
minimise the effect of changes of the above parameters on $N_{\rm
exp}(x)$. We have checked the success of this strategy by creating six
more sets of simulations, each consisting of $10$ datasets ($= 100$
spectra). For each of these sets we varied one parameter: $\beta =
1.7$, $\beta = 1.3$, $B$ (overall normalisation) decreased and
increased by $20$ per cent, $\mu_b = 40$~km~s$^{-1}$, and S/N $=
10$. We have found no significant variation of the total number of LM
in any of these simulations. The shapes of the distributions of LM as
functions of $RG$ and FWHM$_{\rm s}$ also agree well with the
distributions for the `standard' case. The only exception was the
$\beta = 1.3$ model which produces less significant LM compared to the
`standard' case (although this has little effect on $\zeta(RG)$). In
assessing the robustness of the result of Fig.~\ref{lm_s} with respect
to our choice of parameter values we thus only need to consider how
the individual parameters affect $N_{\rm obs}$.

Given equation \eref{norm}, $\sigma_G$ is virtually independent of
$\gamma$ for all reasonable values. Thus the only effect of changing
$\gamma$ is to change the size of the fluctuations, $G - \langle G
\rangle$ in equation \eref{rg}.  Because we normalise $\langle G
\rangle$ at the central wavelength of a spectrum (see equation
\ref{norm}) an increase in $\gamma$ will decrease the magnitude of
overdensities at longer wavelengths while increasing the magnitude of
overdensities at shorter wavelengths and vice versa. Thus there will
only be an appreciable effect if overdensities tend to lie to one side
of a spectrum which is not the case. In addition, a wrong $\gamma$
value cannot explain why the local minima occur almost exclusively on
certain velocity scales.  Nevertheless, we re-determined the LM for
the real data for $\gamma = 2.2$ and $\gamma = 2.8$ and the effect is
small.

By determining the normalisation of the effective optical depth for
each spectrum individually we avoid over- or underestimation of $G -
\langle G \rangle$ caused by small errors in the zeroth and first
order of the continuum fits of the {\em individual} spectra. However,
if the continua are {\em systematically} low or high then we will
under- or overestimate the mean optical depth normalisation, $B$, and
thus under- or overestimate $\sigma_G$.  However, re-analysing the
real data with $B$ increased by $20$ per cent we still find $98$ LM,
significantly above the expected number of $51\pm7.6$.

The parameters $\mu_b$, $\sigma_b$, and $b_{\rm cut}$ all affect the
parameter $q$ in equation \eref{sigG}. \citeN{Kim97} reported a
possible evolution of $\mu_b$ with redshift over a range covered by
the present data. If this were correct and if too small a value for
$q$ were indeed the reason for the observed excess of LM then one
would expect the excess to be larger at smaller redshifts where
$\mu_b$ may be higher. Separating the data into a high and a low
redshift bin we find $N_{\rm obs} (N_{\rm exp}) = 53 (25\pm5.0)$ and
$50 (26\pm5.2)$ for the low and high redshift bins respectively. In
any case, we have re-analysed the data for $\mu_b = 40$~km~s$^{-1}$
(which corresponds to an increase of $q$ of $\sim 30$ per cent, since
$q$ is roughly linear in $\mu_b$) and still found $84$ LM.

Finally we have considered $\beta = 1.3$ which also has the effect of
increasing $\sigma_G$. A re-analysis of the data yielded $79$ LM which
is still too large to be compatible with the simulations.

From the tests described above we conclude that the result presented
in Fig.~\ref{lm_s} is quite robust: there exist structures in the
\lya\ forest on scales of up to $\sim 1200$~km~s$^{-1}$ at the $> 99$
per cent confidence level. This result constitutes clear evidence for
the non-uniformity of the \lya\ forest on large scales despite
repeated findings that the two-point correlation function of
absorption lines shows no signal on scales $\ga 300$~km~s$^{-1}$
(e.g. \citeNP{Cristiani97}).

\subsection{Double lines of sight} \label{dlos}
\begin{table*}
\caption{Pairs of QSOs grouped according to their proper transverse
separation, $d$.}
\label{grouptab}
\begin{tabular}{ccc}
\hline
$d < 3$~h$^{-1}$~Mpc & $3$~h$^{-1}$~Mpc $< d < 6$~h$^{-1}$~Mpc &
$6$~h$^{-1}$~Mpc $< d$ \\
\hline
Q0041-2638 Q0042-2639 & Q0041-2638 Q0042-2627 & Q0041-2638 Q0041-2707 \\
Q0041-2707 Q0041-2658 &	Q0041-2638 Q0042-2656 &	Q0041-2638 Q0041-2607 \\
Q0042-2627 Q0042-2639 &	Q0041-2638 Q0043-2633 &	Q0041-2638 Q0042-2657 \\
Q0042-2627 Q0043-2633 &	Q0041-2707 Q0042-2714 &	Q0041-2707 Q0041-2607 \\
Q0042-2639 Q0043-2633 &	Q0041-2707 Q0042-2656 &	Q0041-2707 Q0042-2627 \\
Q0042-2656 Q0042-2657 &	Q0041-2707 Q0042-2657 &	Q0041-2607 Q0041-2658 \\
		      &	Q0041-2607 Q0042-2627 &	Q0041-2607 Q0042-2714 \\
		      &	Q0041-2658 Q0042-2714 &	Q0041-2607 Q0042-2657 \\
		      &	Q0041-2658 Q0042-2657 &	Q0042-2627 Q0042-2656 \\
		      &	Q0042-2639 Q0042-2656 &	Q0042-2627 Q0042-2657 \\
		      &	Q0042-2639 Q0042-2657 & \\
	              & Q0042-2714 Q0042-2657 & \\
		      &	Q0042-2656 Q0043-2633 & \\
		      &	Q0042-2657 Q0043-2633 & \\
\hline
$\overline{d} = 2.45$~h$^{-1}$~Mpc & $\overline{d} = 4.59$~h$^{-1}$~Mpc &
$\overline{d} = 9.67$~h$^{-1}$~Mpc \\
\hline
\end{tabular}
\end{table*}

So far we have not taken advantage of the fact that the QSOs of Table
\ref{qsotab} are a close group in the plane of the sky. We shall now
examine possible correlations {\em across} lines of sight. One of the
advantages of the analysis used above is that it is easily applied to
multiple lines of sight: transmission triangles are simply averaged
where they overlap. Given the transverse separations of the QSOs in
this group and assuming reasonable absorber sizes, two different lines
of sight will not probe the same absorbers. According to our
null-hypothesis of an unclustered \lya\ forest, two different lines of
sight are therefore uncorrelated. Thus the variance of a mean
transmission triangle (averaged over multiple lines of sight) is given
by $\sigma_G^2(\lambda, \sigma_{\rm s}) / n$, where $n$ is the number
of triangles used at $(\lambda, \sigma_{\rm s})$. This procedure
should enhance structures that extend across multiple lines of sight
and surpress those that do not.

Here we present our results for the case $n = 2$. We have sorted all
pairs of QSOs into one of three groups according to the proper
transverse separation, $d$, of the pair (calculated at the redshift of
the lower redshift QSO): $d < 3$~h$^{-1}$~Mpc, $3$~h$^{-1}$~Mpc $< d <
6$~h$^{-1}$~Mpc and $6$~h$^{-1}$~Mpc $< d$. Table \ref{grouptab}
lists those pairs within the groups for which the relevant spectral
regions overlap.

\begin{figure*}
\psfig{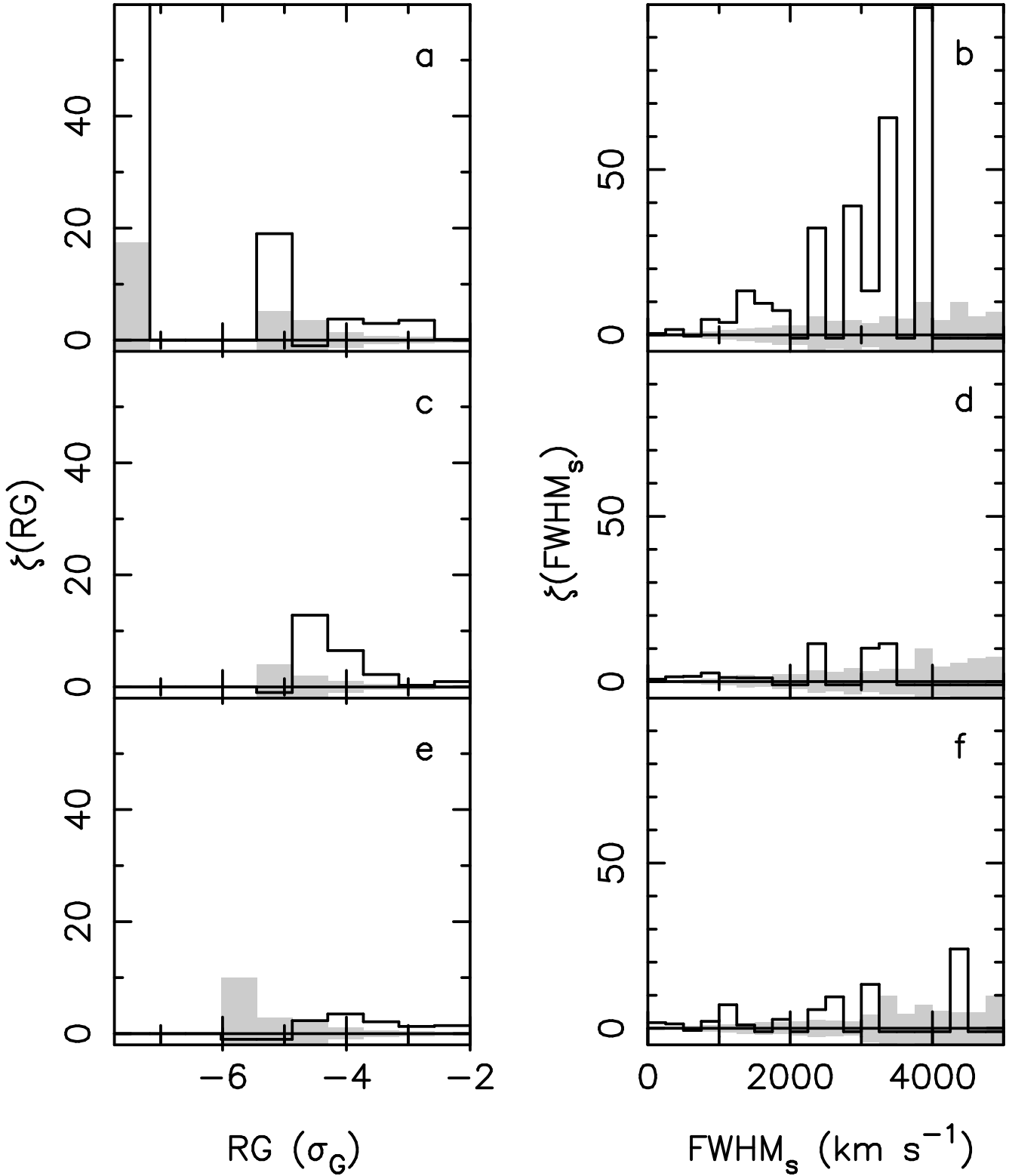}
\caption{Excess of observed number of local minima in reduced mean
transmission triangles of QSO pairs over the expected number ($=$ mean
of $100$ simulations) as a function of the significance level (panels
a, c, e) and the smoothing scale (panels b, d, f). The value of the
leftmost bin of panel (a) is $303$. $\pm1$~rms levels as derived from
the simulations are indicated by the grey shaded areas. The top row
shows the result for QSO pairs with $d < 3$~h$^{-1}$~Mpc, the middle
row for $3$~h$^{-1}$~Mpc $< d < 6$~h$^{-1}$~Mpc and the bottom row for
$6$~h$^{-1}$~Mpc $< d$.}
\label{lm_d}
\end{figure*}

As in Section \ref{lm} we proceeded to construct $\zeta$, but this
time using only the overlap regions of the reduced mean triangles.
Fig.~\ref{lm_d} shows $\zeta(RG)$ and $\zeta({\rm FWHM}_{\rm s})$ for
the three cases listed above. The top row shows the results for small
transverse separations, the middle row for intermediate, and the
bottom row for large transverse separations. There is evidence for
a trend: with increasing separation we detect fewer and fewer
overdensities at large significance levels and large smoothing scales
relative to the simulations.

In panel (a) there is an excess of LM which is somewhat weaker than
that seen for single lines of sight (Fig.~\ref{lm_s}a). However, we
detected one LM at $-7.5\sigma_G$. In contrast, the $100$ simulated
datasets revealed {\em no} LM with $RG < -6\sigma_G$. (Extrapolating
from lower significance levels gives $N_{\rm exp} = 0.0033$ for this
bin.) This very significant overdensity lies at $z = 2.272$ and is
produced by the near-coincidence in redshift space ($\Delta v \approx
450$~km~s$^{-1}$) of the two most significant single-line of sight
overdensities of the entire sample. These two overdensities are found
in the spectra of Q0041--2658 and Q0041--2707. The two lines of sight
are separated by $2.31$~h$^{-1}$~Mpc and are the second closest pair
in the sample. This is a very clear example of a coherent structure
traced by \lya\ absorption extending across two lines of sight.

In all we detect $N_{\rm obs} (N_{\rm exp}) = 41 (14\pm4.7)$, $62
(28\pm6.9)$, and $65 (25\pm6.4)$ LM for small, intermediate, and large
separations respectively. Thus the excess of the {\em total} number of
LM does not seem to vary. However, the {\em shapes} of the
distributions change quite significantly with line of sight
separation. In Table \ref{ks} we list KS probabilities that the
simulated and observed distributions of $RG$ values and smoothing
scales agree for the three groups.
\begin{table}
\caption{Kolmogorov--Smirnov probabilities that the observed and
expected distributions of $RG$ values and smoothing scales agree for
the three line of sight separation groups.}
\label{ks}
\begin{tabular}{rcc}
\hline
& P$_{\rm KS}(RG)$ & P$_{\rm KS}($FWHM$_{\rm s})$\\
\hline
$d < 3$:     & $7 \times 10^{-5}$ & $8 \times 10^{-5}$\\
$3 < d < 6$: & $0.011$            & $0.017$\\
$6 < d$:     & $0.49$             & $0.46$\\
\hline
\end{tabular}
\end{table}
At small separations, the simulated distributions disagree strongly
with the observations, producing too few LM at large significance
levels and at large smoothing scales. However at large separations,
the distributions agree very well.

If the observed excess of LM in Figs~\ref{lm_s} and \ref{lm_d}(a) were
simply due to an incorrect choice of the values for the parameters of
equation \eref{Vd} as discussed in Section \ref{lm}, then it is hard
to understand why that excess should be so strongly reduced for large
line of sight separations. The removal of the pair Q0041--2707 -
Q0041--2658 leaves the result qualitatively unchanged as does the
removal of substructure as outlined in Section \ref{lm}. The removal
of all LM that have a metal system within $300$~km~s$^{-1}$ in either
of the spectra of the pair somewhat weakens the excess of panel (a)
but does not remove the significance of the effect, since we still
obtain P$_{\rm KS}(RG) = 9 \times 10^{-4}$ and P$_{\rm KS}($FWHM$_{\rm
s}) = 3 \times 10^{-5}$ at small separations but good agreement at
large separations (cf. Table \ref{ks}).

A possible explanation for the trend of the decreasing excess with
line of sight separation could be that the group of close QSO pairs is
dominated by those QSOs whose spectra show the most significant
overdensities and that these QSOs are absent from the group of
large-separation pairs. However, by inspection of Table \ref{grouptab}
we can see that each of the groups contains at least eight of the ten
QSOs at least once. In particular, Q0041--2707 and Q0041--2658 are
present in all groups and both appear in the group of close QSOs only
once.

\begin{figure}
\vspace{1cm}
\psfig{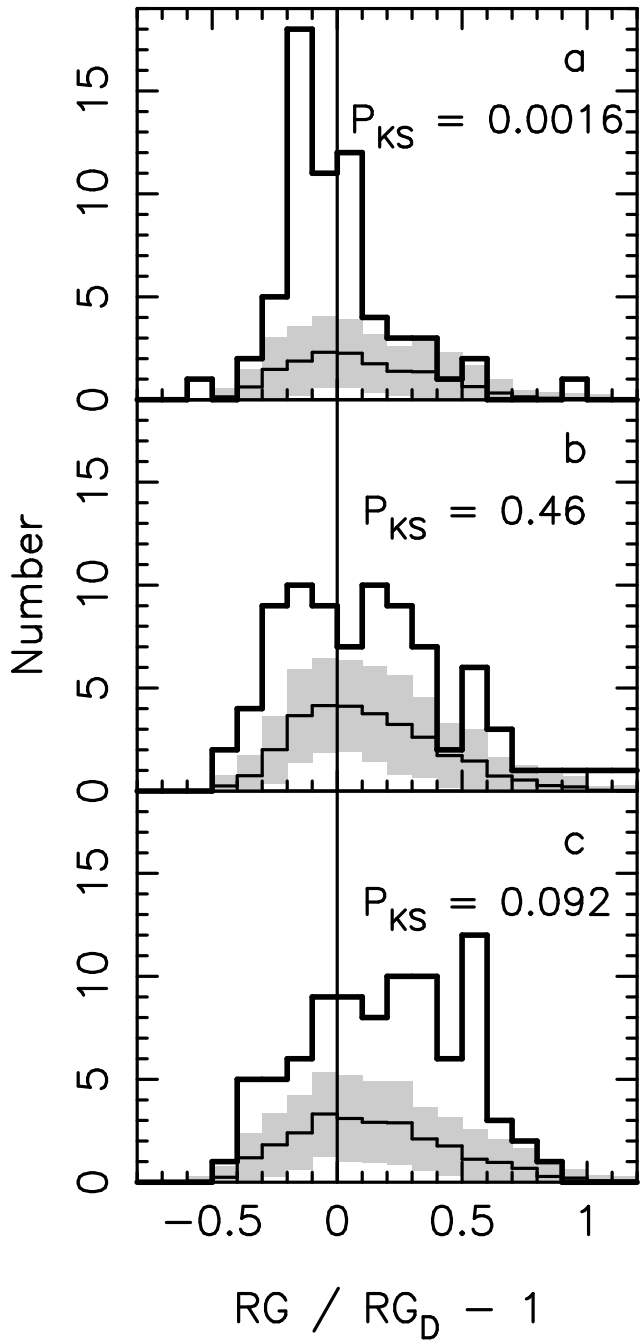}
\caption{Histograms of the indicated quantity for all local minima
detected in the reduced mean transmission triangles of QSO pairs
(thick lines). The thin lines show the mean histograms computed from
$100$ simulations. $\pm1$~rms levels as derived from the simulations
are indicated by the grey shaded areas. The top row shows the result
for QSO pairs with $d < 3$~h$^{-1}$~Mpc, the middle row for
$3$~h$^{-1}$~Mpc $< d < 6$~h$^{-1}$~Mpc and the bottom row for
$6$~h$^{-1}$~Mpc $< d$. We also show KS probabilities that the
simulated and observed distributions agree.}
\label{slm_dlm}
\end{figure}

To further investigate whether the excess of LM seen in the double
lines of sight (DLOS) is simply due to the excess already detected in
the single lines of sight (SLOS) we attempt to identify the structures
in the SLOS that give rise to the LM detected in the DLOS: for each LM
in a DLOS we search for the most significant LM in each of the
constituent SLOS within $1000$~km~s$^{-1}$ (our results do not depend
sensitively on this value). If the LM in the DLOS is due to a
structure extending across the two lines of sight then one would
expect less significant LM in both of the constituent SLOS. However,
if the LM in the DLOS is due to a strong overdensity which does not
extend across both SLOS then one would expect to find a LM in only one
of the SLOS. If LM are found in both SLOS, then they may be quite
dissimilar and one should be of greater significance than the LM in
the DLOS. For each LM in the DLOS we have computed the following
quantity:
\[
\frac{RG_i}{RG_{\rm D}} - 1,  \quad \quad i = 1,2
\]
where $RG_i$ denotes the LM in the two SLOS and $RG_{\rm D}$ that of
the DLOS. This quantity measures how the SLOS-LM compare with the
DLOS-LM they produce. In Fig.~\ref{slm_dlm} we plot as thick solid
lines the histograms of this measure for the three groups of QSO pairs
(top = small separation, bottom = large separation). We have also
computed the same histograms for the $100$ simulated datasets and
display the mean and mean~$\pm$~$1$~rms histograms as thin solid lines
and grey shaded areas respectively. There is a small number of cases
where no LM can be found in either of the single sight-lines and these
are excluded from this analysis.

All panels of Fig.~\ref{slm_dlm} show a large excess of the observed
distributions over the simulated ones. This is not overly surprising
since we already know that the total number of LM exceeds the expected
number at all separations. However, at small separations (panel a) the
excess is skewed to values $RG / RG_{\rm D} - 1 < 0$. A KS test shows
that the simulated and observed distributions disagree at the $> 99$
per cent confidence level. This discrepancy disappears for larger line
of sight separations.

In addition, we have also counted the number of DLOS-LM which are
produced by only a single SLOS-LM and found that the excess of such
cases increases from $2.2\sigma$ over $3.4\sigma$ to $4.4\sigma$ for
increasing line of sight separation.

Again, the removal of certain subclasses of LM (those of the
Q0041--2707 - Q0041--2658 pair, substructure, those associated with
metal lines) does not change the results significantly.

In summary, the results above show that for small line of sight
separations two SLOS-LM combine to make a DLOS-LM of greater
significance, whereas for large line of sight separations the
structures seen in the SLOS are `diluted'. We thus take the
anti-correlation of the excess of LM seen in the reduced mean
transmission triangles of double-lines of sight with sight-line
separation as strong evidence for the existence of structures on
scales of up to $3$~h$^{-1}$ proper Mpc in the \lya\ forest.

\subsection{Correlation with metal lines} \label{cml}
Not surprisingly, all the metal systems that were found in the present
data have associated overdense \lya\ absorption. We have repeatedly
remarked in the sections above that the removal of LM that are
associated with metal lines does not change the results
qualitatively. Here, we take the opposite approach: we have repeated
the analysis of the previous section only for those DLOS-LM that have
a metal system within $1000$~km~s$^{-1}$ in either of the constituent
SLOS. The corresponding figure to Fig.~\ref{lm_d} shows only a
marginal trend with line of sight separation, with P$_{\rm KS}(RG) =
0.091$ for small separations, which is entirely due to the close pair
Q0041--2658 and Q0041--2707. The corresponding figure to
Fig.~\ref{slm_dlm} and its KS tests also indicate that the DLOS-LM are
due to chance alignments or single SLOS-LM at {\em all} separations,
however the numbers are very low and we cannot draw any definite
conclusions.

It must be kept in mind that the subdivision into LM with and without
metals is subject to a strong selection effect since the work by
\citeN{Cowie95} and \citeN{Songaila96} has clearly shown that the
census of C~{\small IV} in the present data must be significantly
incomplete. Thus the distinction made here is more one between
high-column density vs. low-column density structures rather than
between metals vs. \lya\ only. Therefore in the real data we have
selected LM that are due to high-column density structures whereas in
the simulations we have basically drawn a random sample of LM since in
the simulated data the column density is unrelated to the presence of
metals. 

Nevertheless, it appears that the high-column density structures
traced by metal lines do not produce overdense \lya\ absorption at the
distances probed by this sample, with one notable exception, in
contrast to the lower column density \lya\ only systems. On the other
hand, there is little difference between the distributions of
smoothing scales for those LM that are associated with metal lines and
those that are not. A KS test gives a probability of $0.3$ that the
two distributions are the same.

\subsection{Triple lines of sight} \label{tlos}

\begin{table*}
\caption{Triplets of QSOs grouped according to their proper transverse
separation, $d$.}
\label{group2tab}
\begin{tabular}{ccc}
\hline
$d < 3$~h$^{-1}$~Mpc & $3$~h$^{-1}$~Mpc $< d < 6$~h$^{-1}$~Mpc &
$6$~h$^{-1}$~Mpc $< d$ \\
\hline
Q0042-2627 Q0042-2639 Q0043-2633 & Q0041-2638 Q0042-2656 Q0043-2633 &
Q0041-2638 Q0041-2707 Q0041-2607\\
& Q0041-2707 Q0042-2714 Q0042-2657 & Q0041-2638 Q0041-2607 Q0042-2657\\
& Q0041-2658 Q0042-2714 Q0042-2657 & \\
\hline
$\overline{d} = 2.82$~h$^{-1}$~Mpc & $\overline{d} = 4.73$~h$^{-1}$~Mpc &
$\overline{d} = 8.92$~h$^{-1}$~Mpc \\
\hline
\end{tabular}
\end{table*}

QSO triplets can provide important constraints on the shape of \lya\
absorbers. It is possible that the excess of LM seen in the DLOS is
mainly due to filamentary structures. If this were the case then one
would expect LM of close triple lines of sight (TLOS) to be caused by
two, not three SLOS-LM. However, if the absorption is sheet-like in
nature then one would expect to find three similar SLOS-LM per
TLOS-LM.

We have repeated the analysis of Section \ref{dlos} for triple lines
of sight. Table \ref{group2tab} lists triplets of QSOs grouped
according to their pairwise transverse line of sight separations as in
Table \ref{grouptab}. The triplets of the first two groups form more
or less equilateral triangles on the sky. However, due to the larger
extent of the full group in the DEC direction than in the RA
direction, the triplets of the last group form somewhat flatter
triangles. Unfortunately, our QSO sample is not dense enough to
provide more than one to three triplets per group.

\begin{figure*}
\psfig{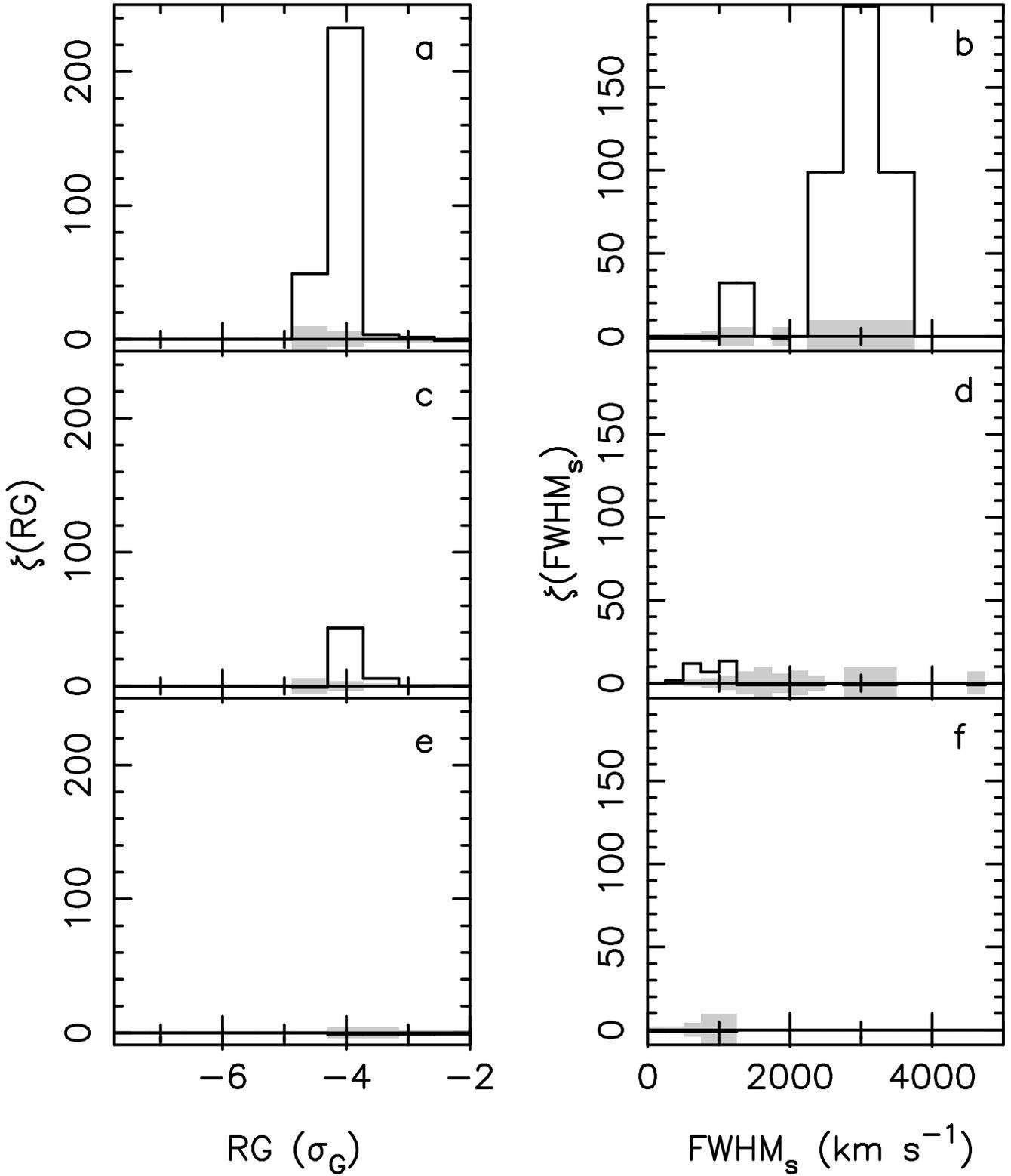}
\caption{Excess of observed number of local minima in reduced mean
transmission triangles of QSO triplets over the expected number ($=$
mean of $100$ simulations) as a function of the significance level
(panels a, c, e) and the smoothing scale (panels b, d, f). $\pm1$~rms
levels as derived from the simulations are indicated by the grey
shaded areas. The top row shows the result for QSO triplets with
pairwise separations $d < 3$~h$^{-1}$~Mpc, the middle row for
$3$~h$^{-1}$~Mpc $< d < 6$~h$^{-1}$~Mpc and the bottom row for
$6$~h$^{-1}$~Mpc $< d$.}
\label{lm_t}
\end{figure*}

Fig.~\ref{lm_t} shows the results. Again we see an anti-correlation of
the excess of LM with line of sight separation (panels a, c, and
e). However, for the smallest separations we find that $7$ of the $10$
identified TLOS-LM are due to only two (and not three) SLOS-LM. The
vast majority of SLOS-LM are of smaller significance than their
respective TLOS-LM. Therefore it seems that at least in the case of
this particular QSO triplet the absorption more often extends across
only two lines of sight than across three. However, the numbers are
small and thus we cannot draw any definite conclusions from this
result.

\section{Conclusions and discussion} \label{conclusions}
We summarise our main results as follows:

(1) We have analysed the \lya\ forest spectra of ten QSOs at $\langle
z \rangle = 2.81$ contained within a $\sim 1$~deg$^2$ field using a
new technique based on the statistics of the transmitted
flux. Comparison with two-point correlation function analyses (Section
\ref{lm}), along with the results of paper I, suggests that this new
method is more sensitive to the presence of large-scale structure than
the two-point correlation function of individually identified
absorption lines.

(2) We find structure on scales of up to $1200$~km~s$^{-1}$ along the
line of sight and on scales of up to $17$~h$^{-1}$~Mpc (comoving) in
the transverse direction. We confirm the existence of large-scale
structure in the \lya\ forest at the $> 99$ per cent confidence level
(\citeNP{Pando96}; \citeNP{Williger99}).

(3) We find strong correlations across lines of sight with proper
separation $< 3$~h$^{-1}$~Mpc. For intermediate separations the
correlation is weaker and there is only little evidence for
correlation at line of sight separations $> 6$~h$^{-1}$~Mpc
(Fig.~\ref{lm_d}). We thus present the first evidence for a dependence
of the correlation strength on line of sight separation, and place an
upper limit of $6$~h$^{-1}$~Mpc on the transverse correlation scale at
$z = 2.81$.

Assuming that the absorbing structures are expanding with the Hubble
flow, we find that the line of sight and transverse correlation scales
are roughly comparable ($1200$~km~s$^{-1} H^{-1}(z = 2.81) =
1.6$~h$^{-1}$~Mpc) with a suggestion that the absorbers might be
flattened in the line of sight direction since we still detect
somewhat significant correlations on transverse scales of
$4.6$~h$^{-1}$~Mpc. Furthermore, the analysis of the only close QSO
triplet of the sample showed that many coincident absorption features
are common to only two spectra, perhaps indicating an elongated
shape in the plane of the sky. However, no firm conclusions can be
drawn here until more QSO triplets have been analysed.

Using a comparatively small sample with significantly smaller line of
sight separations, \citeN{Crotts89} and \citeN{Crotts98} found a
stronger correlation signal for lines with $W > 0.4$~\AA\ than for
weaker lines. In contrast, \citeN{Williger99} found from their
analysis of the present SGP data that the inclusion of weak lines ($W
> 0.1$~\AA) strengthened their correlation signal. Here we can
tentatively confirm the \citeN{Williger99} result. This is in
agreement with the results of \citeN{CenSim97} who predicted from
their numerical simulations that high-column density lines have
smaller correlation lengths than low-column density ones. This is
already evident from a visual inspection of the three-dimensional
distribution of the absorbing gas in the simulations at different
overdensities. Large overdensities are confined to relatively small,
more or less spherical regions whereas small overdensities form
extended filaments and sheets.

It is interesting to compare our results with predictions from
hydrodynamical simulations. The typical length of the low-column
density filaments in the simulations is of the order of
$1$~h$^{-1}$~Mpc in proper units (\citeNP{Miralda96};
\citeNP{Zhang98}). \citeN{CenSim97} performed a detailed analysis of
double-lines of sight in a $\Lambda$CDM simulation. They concluded
that for proper line of sight separations $> 500$~h$^{-1}$~kpc any
coincident absorption is due to unrelated and spatially uncorrelated
clouds. They pointed out however that a significant amount of power is
missing from their simulation on the scale of the simulated box size
($10$~h$^{-1}$~Mpc). Even after correcting for this effect though, the
transverse correlation scale predicted from these simulations remains
significantly smaller (by about a factor of $3$) than the one derived
in this work. To investigate this discrepancy it will be necessary to
perform a detailed comparison by subjecting simulated spectra, drawn
from a suitable (i.e. large) simulation box, to the same analysis we
have presented here.

\citeN{Croft98} and \citeN{Nusser99} have outlined schemes for
recovering the power spectrum of mass fluctuations from \lya\ forest
spectra. Recently, \citeN{Croft98b} performed the first such
measurement on scales of $2$--$12$~h$^{-1}$~Mpc from $19$
intermediate-resolution QSO spectra at $z = 2.5$. Although we have
concentrated in this work only on identifying typical correlation
scales, our results confirm the usefulness of intermediate-resolution
data for large-scale structure studies when analysing the distribution
of the transmitted flux directly. Since the \lya\ forest is thought to
trace the mass distribution more closely than galaxies we are likely
to gain the most direct measurement of the bias between galaxy and
mass clustering by comparing the power spectrum of the \lya\ forest
with that of galaxies \cite{Croft98b}.

\section*{Acknowledgments}
We thank J.~Baldwin, C.~Hazard, R.~McMahon, and A.~Smette for kindly
providing access to the data. We also thank C.~Lineweaver for helpful
comments on the manuscript. JL acknowledges support from the German
Academic Exchange Service (DAAD) in the form of a PhD scholarship
(Hochschulsonderprogramm III).

\label{lastpage}


\begin{thebibliography}{}

\bibitem[\protect\citeauthoryear{Bahcall et~al.}{Bahcall
  et~al.}{1991}]{Bahcall91}
Bahcall J.~N., Jannuzi B.~T., Schneider D.~P., Hartig G.~F., Bohlin R.,
  Junkkarinen V., 1991, ApJ, 377, L5

\bibitem[\protect\citeauthoryear{Bi \& Davidsen}{Bi \& Davidsen}{1997}]{Bi97}
Bi~H.,  Davidsen A.~F., 1997, ApJ, 479, 523

\bibitem[\protect\citeauthoryear{Bowen, Blades, \& Pettini}{Bowen
  et~al.}{1996}]{Bowen96}
Bowen D.~V., Blades J.~C.,  Pettini M., 1996, ApJ, 464, 141

\bibitem[\protect\citeauthoryear{Carswell et~al.}{Carswell
  et~al.}{1984}]{Carswell84}
Carswell R.~F., Morton D.~C., Smith M.~G., Stockton A.~N., Turnshek D.~A.,
  Weymann R.~J., 1984, ApJ, 278, 486

\bibitem[\protect\citeauthoryear{Cen et~al.}{Cen et~al.}{1994}]{Cen94}
Cen R., Miralda-Escud\'e J., Ostriker J.~P.,  Rauch M., 1994, ApJ, 437, L9

\bibitem[\protect\citeauthoryear{Cen \& Simcoe}{Cen \& Simcoe}{1997}]{CenSim97}
Cen R.,  Simcoe R.~A., 1997, ApJ, 483, 8

\bibitem[\protect\citeauthoryear{Charlton, Churchill, \& Linder}{Charlton
  et~al.}{1995}]{Charlton95}
Charlton J.~C., Churchill C.~W.,  Linder S.~M., 1995, ApJ, 452, L81

\bibitem[\protect\citeauthoryear{Chen et~al.}{Chen et~al.}{1998}]{Chen98}
Chen H.-W., Lanzetta K.~M., Webb J.~K.,  Barcons X., 1998, ApJ, 498, 77

\bibitem[\protect\citeauthoryear{Chernomordik}{Chernomordik}{1995}]{Chernomord%
ik95}
Chernomordik V.~V., 1995, ApJ, 440, 431

\bibitem[\protect\citeauthoryear{Cowie et~al.}{Cowie et~al.}{1995}]{Cowie95}
Cowie L.~L., Songaila A., Kim T.-S.,  Hu~E.~M., 1995, AJ, 109, 1522

\bibitem[\protect\citeauthoryear{Cristiani et~al.}{Cristiani
  et~al.}{1997}]{Cristiani97}
Cristiani S., D'Odorico S., D'Odorico V., Fontana A., Giallongo E.,  Savaglio
  S., 1997, MNRAS, 285, 209

\bibitem[\protect\citeauthoryear{Cristiani et~al.}{Cristiani
  et~al.}{1995}]{Cristiani95b}
Cristiani S., D'Odorico S., Fontana A., Giallongo E.,  Savaglio S., 1995,
  MNRAS, 273, 1016

\bibitem[\protect\citeauthoryear{Croft et~al.}{Croft et~al.}{1998a}]{Croft98}
Croft R.~A.~C., Weinberg D.~H., Katz N.,  Hernquist L., 1998a, ApJ, 495, 44

\bibitem[\protect\citeauthoryear{Croft et~al.}{Croft et~al.}{1998b}]{Croft98b}
Croft R.~A.~C., Weinberg D.~H., Pettini M., Hernquist L.,  Katz N., 1998b, ApJ,
  submitted, astro-ph/9809401

\bibitem[\protect\citeauthoryear{Crotts}{Crotts}{1989}]{Crotts89}
Crotts A.~P.~S., 1989, ApJ, 336, 550

\bibitem[\protect\citeauthoryear{Crotts \& Fang}{Crotts \&
  Fang}{1998}]{Crotts98}
Crotts A.~P.~S.,  Fang Y., 1998, ApJ, 502, 16

\bibitem[\protect\citeauthoryear{Dav\'e et~al.}{Dav\'e et~al.}{1998a}]{Dave98}
Dav\'e R., Hellsten U., Hernquist L., Katz N.,  Weinberg D.~H., 1998a, ApJ,
  509, 661

\bibitem[\protect\citeauthoryear{Dav\'e et~al.}{Dav\'e et~al.}{1998b}]{Dave98b}
Dav\'e R., Hernquist L., Katz N.,  Weinberg D.~H., 1998b, ApJ, 511, 521

\bibitem[\protect\citeauthoryear{Dav\'e et~al.}{Dav\'e et~al.}{1997}]{Dave97}
Dav\'e R., Hernquist L., Weinberg D.~H.,  Katz N., 1997, ApJ, 477, 21

\bibitem[\protect\citeauthoryear{Dinshaw et~al.}{Dinshaw
  et~al.}{1998}]{Dinshaw98}
Dinshaw N., Foltz C.~B., Impey C.~D.,  Weyman R.~J., 1998, ApJ, 494, 567

\bibitem[\protect\citeauthoryear{Dinshaw et~al.}{Dinshaw
  et~al.}{1997}]{Dinshaw97}
Dinshaw N., Weyman R.~J., Impey C.~D., Foltz C.~B., Morris S.~L.,  Ake T.,
  1997, ApJ, 491, 45

\bibitem[\protect\citeauthoryear{D'Odorico et~al.}{D'Odorico
  et~al.}{1998}]{DOdorico98}
D'Odorico V., Cristiani S., D'Odorico S., Fontana A., Giallongo E.,  Shaver P.,
  1998, A\&A, 339, 678

\bibitem[\protect\citeauthoryear{Elowitz, Green, \& Impey}{Elowitz
  et~al.}{1995}]{Elowitz95}
Elowitz R.~M., Green R.~F.,  Impey C.~D., 1995, ApJ, 440, 458

\bibitem[\protect\citeauthoryear{Fang et~al.}{Fang et~al.}{1996}]{Fang96}
Fang Y., Duncan R.~C., Crotts A.~P.~S.,  Bechtold J., 1996, ApJ, 462, 77

\bibitem[\protect\citeauthoryear{Fern\'andez-Soto et~al.}{Fern\'andez-Soto
  et~al.}{1996}]{Fernandez96}
Fern\'andez-Soto A., Lanzetta K.~M., Barcons X., Carswell R.~F., Webb J.~K.,
  Yahil A., 1996, ApJ, 460, L85

\bibitem[\protect\citeauthoryear{Gnedin}{Gnedin}{1998}]{Gnedin98}
Gnedin N.~Y., 1998, MNRAS, 294, 407

\bibitem[\protect\citeauthoryear{Gnedin \& Hui}{Gnedin \&
  Hui}{1998}]{Gnedin98b}
Gnedin N.~Y.,  Hui L., 1998, MNRAS, 296, 44

\bibitem[\protect\citeauthoryear{Hernquist et~al.}{Hernquist
  et~al.}{1996}]{Hernquist96}
Hernquist L., Katz N., Weinberg D.~H.,  Miralda-Escud\'e J., 1996, ApJ, 457,
  L51

\bibitem[\protect\citeauthoryear{Hu et~al.}{Hu et~al.}{1995}]{Hu95}
Hu~E.~M., Kim T.-S., Cowie L.~L., Songaila A.,  Rauch M., 1995, AJ, 110, 1526

\bibitem[\protect\citeauthoryear{Hui, Gnedin, \& Zhang}{Hui
  et~al.}{1997}]{Hui97}
Hui L., Gnedin N.~Y.,  Zhang Y., 1997, ApJ, 486, 599

\bibitem[\protect\citeauthoryear{Ikeuchi}{Ikeuchi}{1986}]{Ikeuchi86b}
Ikeuchi S., 1986, ApJS, 118, 509

\bibitem[\protect\citeauthoryear{Ikeuchi \& Ostriker}{Ikeuchi \&
  Ostriker}{1986}]{Ikeuchi86}
Ikeuchi S.,  Ostriker J.~P., 1986, ApJ, 301, 522

\bibitem[\protect\citeauthoryear{Khare et~al.}{Khare et~al.}{1997}]{Khare97}
Khare P., Srianand R., York D.~G., Green R., Welty D., Huang K.-L.,  Bechtold
  J., 1997, MNRAS, 285, 167

\bibitem[\protect\citeauthoryear{Kim et~al.}{Kim et~al.}{1997}]{Kim97}
Kim T.-S., Hu~E.~M., Cowie L.~L.,  Songaila A., 1997, AJ, 114, 1

\bibitem[\protect\citeauthoryear{Kirkman \& Tytler}{Kirkman \&
  Tytler}{1997}]{Kirkman97}
Kirkman D.,  Tytler D., 1997, ApJ, 484, 672

\bibitem[\protect\citeauthoryear{Kulkarni et~al.}{Kulkarni
  et~al.}{1996}]{Kulkarni96}
Kulkarni V.~P., Huang K., Green R.~F., Bechtold J., Welty D.~E.,  York D.~G.,
  1996, MNRAS, 279, 197

\bibitem[\protect\citeauthoryear{Lanzetta et~al.}{Lanzetta
  et~al.}{1995}]{Lanzetta95}
Lanzetta K.~M., Bowen D.~B., Tytler D.,  Webb J.~K., 1995, ApJ, 442, 538

\bibitem[\protect\citeauthoryear{Le~Brun \& Bergeron}{Le~Brun \&
  Bergeron}{1998}]{LeBrun98}
Le~Brun V.,  Bergeron J., 1998, A\&A, 332, 814

\bibitem[\protect\citeauthoryear{Le~Brun, Bergeron, \& Boiss\'e}{Le~Brun
  et~al.}{1996}]{LeBrun96}
Le~Brun V., Bergeron J.,  Boiss\'e P., 1996, A\&A, 306, 691

\bibitem[\protect\citeauthoryear{Liske, Webb, \& Carswell}{Liske
  et~al.}{1998}]{Liske98b}
Liske J., Webb J.~K.,  Carswell R.~F., 1998, MNRAS, 301, 787

\bibitem[\protect\citeauthoryear{Lu et~al.}{Lu et~al.}{1998}]{Lu98}
Lu~L., Sargent W.~L.~W., Barlow T.~A.,  Rauch M., 1998, AJ, submitted,
  astro-ph/9802189

\bibitem[\protect\citeauthoryear{Lu et~al.}{Lu et~al.}{1996}]{Lu96}
Lu~L., Sargent W.~L.~W., Womble D.~S.,  Takada-Hidai M., 1996, ApJ, 472, 509

\bibitem[\protect\citeauthoryear{Meiksin \& Bouchet}{Meiksin \&
  Bouchet}{1996}]{Meiksin95}
Meiksin A.,  Bouchet F.~R., 1996, MNRAS, 283, 1388

\bibitem[\protect\citeauthoryear{Miralda-Escud\'e et~al.}{Miralda-Escud\'e
  et~al.}{1996}]{Miralda96}
Miralda-Escud\'e J., Cen R., Ostriker J.~P.,  Rauch M., 1996, ApJ, 471, 582

\bibitem[\protect\citeauthoryear{Miralda-Escud\'e \& Rees}{Miralda-Escud\'e \&
  Rees}{1997}]{Miralda97}
Miralda-Escud\'e J.,  Rees M.~J., 1997, ApJ, 478, L57

\bibitem[\protect\citeauthoryear{Mo et~al.}{Mo et~al.}{1992}]{Mo92}
Mo~H.~J., Xia X.~Y., Deng Z.~G., B{\"o}rner G.,  Fang L.~Z., 1992, A\&A, 256,
  L23

\bibitem[\protect\citeauthoryear{Morris et~al.}{Morris et~al.}{1991}]{Morris91}
Morris S.~L., Weymann R.~J., Savage B.~D.,  Gilliland R.~L., 1991, ApJ, 377,
  L21

\bibitem[\protect\citeauthoryear{M{\"u}cket et~al.}{M{\"u}cket
  et~al.}{1996}]{Muecket96}
M{\"u}cket J.~P., Petitjean P., Kates R.~E.,  Riediger R., 1996, A\&A, 308, 17

\bibitem[\protect\citeauthoryear{Nusser \& Haehnelt}{Nusser \&
  Haehnelt}{1999}]{Nusser99}
Nusser A.,  Haehnelt M., 1999, MNRAS, 303, 179

\bibitem[\protect\citeauthoryear{Oke \& Korycansky}{Oke \&
  Korycansky}{1982}]{Oke82}
Oke J.~B.,  Korycansky D.~G., 1982, ApJ, 255, 11

\bibitem[\protect\citeauthoryear{Ostriker, Bajtlik, \& Duncan}{Ostriker
  et~al.}{1988}]{Ostriker88}
Ostriker J.~P., Bajtlik S.,  Duncan R.~C., 1988, ApJ, 327, L35

\bibitem[\protect\citeauthoryear{Ostriker \& Ikeuchi}{Ostriker \&
  Ikeuchi}{1983}]{Ostriker83}
Ostriker J.~P.,  Ikeuchi S., 1983, ApJ, 268, L63

\bibitem[\protect\citeauthoryear{Pando \& Fang}{Pando \& Fang}{1996}]{Pando96}
Pando J.,  Fang L.-Z., 1996, ApJ, 459, 1

\bibitem[\protect\citeauthoryear{Pando \& Fang}{Pando \& Fang}{1998}]{Pando98b}
Pando J.,  Fang L.-Z., 1998, A\&A, 340, 335

\bibitem[\protect\citeauthoryear{Petitjean, M{\"u}cket, \& Kates}{Petitjean
  et~al.}{1995}]{Petitjean95}
Petitjean P., M{\"u}cket J.~P.,  Kates R.~E., 1995, A\&A, 295, L9

\bibitem[\protect\citeauthoryear{Petitjean et~al.}{Petitjean
  et~al.}{1998}]{Petitjean98}
Petitjean P., Surdej J., Smette A., Shaver P., M{\"u}cket J.,  Remy M., 1998,
  A\&A, 334, L45

\bibitem[\protect\citeauthoryear{Rauch \& Haehnelt}{Rauch \&
  Haehnelt}{1995}]{Rauch95}
Rauch M.,  Haehnelt M.~G., 1995, MNRAS, 275, 76

\bibitem[\protect\citeauthoryear{Rees}{Rees}{1986}]{Rees86}
Rees M.~J., 1986, MNRAS, 218, 25

\bibitem[\protect\citeauthoryear{Riediger, Petitjean, \& M{\"u}cket}{Riediger
  et~al.}{1998}]{Riediger98}
Riediger R., Petitjean P.,  M{\"u}cket J.~P., 1998, A\&A, 329, 30

\bibitem[\protect\citeauthoryear{Sargent et~al.}{Sargent et~al.}{1980}]{SYBT}
Sargent W.~L.~W., Young P.~J., Boksenberg A.,  Tytler D., 1980, ApJS, 42, 41

\bibitem[\protect\citeauthoryear{Smette et~al.}{Smette et~al.}{1995}]{Smette95}
Smette A., Robertson J.~G., Shaver P.~A., Reimers D., Wisotzki L.,  K{\"o}hler
  T., 1995, A\&AS, 113, 199

\bibitem[\protect\citeauthoryear{Songaila \& Cowie}{Songaila \&
  Cowie}{1996}]{Songaila96}
Songaila A.,  Cowie L.~L., 1996, AJ, 112, 335

\bibitem[\protect\citeauthoryear{Theuns, Leonard, \& Efstathiou}{Theuns
  et~al.}{1998}]{Theuns98}
Theuns T., Leonard A.,  Efstathiou G., 1998, MNRAS, 297, L49

\bibitem[\protect\citeauthoryear{Theuns et~al.}{Theuns
  et~al.}{1998}]{Theuns98b}
Theuns T., Leonard A., Efstathiou G., Pearce F.~R.,  Thomas P.~A., 1998, MNRAS,
  301, 478

\bibitem[\protect\citeauthoryear{Tripp, Lu, \& Savage}{Tripp
  et~al.}{1998}]{Tripp98}
Tripp T.~M., Lu~L.,  Savage B.~D., 1998, ApJ, 508, 200

\bibitem[\protect\citeauthoryear{Ulmer}{Ulmer}{1996}]{Ulmer96}
Ulmer A., 1996, ApJ, 473, 110

\bibitem[\protect\citeauthoryear{Wadsley \& Bond}{Wadsley \&
  Bond}{1997}]{Wadsley97}
Wadsley J.~W.,  Bond J.~R., 1997, in ASP Conference Series, Vol. 123, Clarke
  D.~A.,  West M.~J., ed, Proceedings of the 12th Kingston Meeting
  ``Computational Astrophysics'', San Francisco, p. 332, astro-ph/9612148

\bibitem[\protect\citeauthoryear{Webb}{Webb}{1987}]{Webb87a}
Webb J.~K., 1987, in Hewitt A., Burbidge G.,  Fang L.-Z., ed, Proceedings of
  the 124th IAU Symposium ``Observational Cosmology''.
\newblock Reidel, Dordrecht, p. 803

\bibitem[\protect\citeauthoryear{Webb et~al.}{Webb et~al.}{1992}]{Webb92}
Webb J.~K., Barcons X., Carswell R.~F.,  Parnell H.~C., 1992, MNRAS, 255, 319

\bibitem[\protect\citeauthoryear{Williger \& Babul}{Williger \&
  Babul}{1992}]{Williger92}
Williger G.~M.,  Babul A., 1992, ApJ, 399, 385

\bibitem[\protect\citeauthoryear{Williger et~al.}{Williger
  et~al.}{1996}]{Williger96}
Williger G.~M., Hazard C., Baldwin J.~A.,  McMahon R.~G., 1996, ApJS, 104, 145

\bibitem[\protect\citeauthoryear{Williger et~al.}{Williger
  et~al.}{1999}]{Williger99}
Williger G.~M., Smette A., Hazard C., Baldwin J.~A.,  McMahon R.~G., 1999, ApJ,
  in press, astro-ph/9910369

\bibitem[\protect\citeauthoryear{York et~al.}{York et~al.}{1991}]{York91}
York D.~G., Yanny B., Crotts A., Carilli C.,  Garrison E., 1991, MNRAS, 250, 24

\bibitem[\protect\citeauthoryear{Zhang, Anninos, \& Norman}{Zhang
  et~al.}{1995}]{Zhang95}
Zhang Y., Anninos P.,  Norman M.~L., 1995, ApJ, 453, L57

\bibitem[\protect\citeauthoryear{Zhang et~al.}{Zhang et~al.}{1997}]{Zhang97}
Zhang Y., Anninos P., Norman M.~L.,  Meiksin A., 1997, ApJ, 485, 496

\bibitem[\protect\citeauthoryear{Zhang et~al.}{Zhang et~al.}{1998}]{Zhang98}
Zhang Y., Meiksin A., Anninos P.,  Norman M.~L., 1998, ApJ, 495, 63

\end{thebibliography}
\end{document}